\documentclass[%
 reprint,
superscriptaddress,
 amsmath,amssymb,
 aps,
prb,
floatfix,
]{revtex4-2}

\usepackage{CJK}
\usepackage{amsmath, amssymb}
\usepackage{graphicx}
\usepackage{dcolumn}
\usepackage{bm}
\usepackage{color}

\begin{document}

\preprint{APS/123-QED}

\title{Exciting space-time surface plasmon polaritons by irradiating a nanoslit structure}
\author{Naoki Ichiji}
\affiliation{Graduate School of Pure and Applied Sciences, University of Tsukuba, 1-1-1 Tennodai, Tsukuba-shi, Ibaraki 305-8571, Japan}
\author{Murat Yessenov}
\affiliation{CREOL, The College of Optics \& Photonics, University of Central Florida, Orlando, FL 32816, USA}
\author{Kenneth L. Schepler}
\affiliation{CREOL, The College of Optics \& Photonics, University of Central Florida, Orlando, FL 32816, USA}
\author{Ayman F. Abouraddy}
\thanks{raddy@creol.ucf.edu}
\affiliation{CREOL, The College of Optics \& Photonics, University of Central Florida, Orlando, FL 32816, USA}
\author{Atsushi Kubo}
\email{kubo.atsushi.ka@u.tsukuba.ac.jp}
\affiliation{Faculty of Pure and Applied Sciences, University of Tsukuba,1-1-1 Tennodai, Tsukuba-shi, Ibaraki 305-8571, Japan}

\date{\today}

\begin{abstract}
Space-time (ST) wave packets are propagation-invariant pulsed optical beams that travel freely in dielectrics at a tunable group velocity without diffraction or dispersion. Because ST wave packets maintain these characteristics even when only one transverse dimension is considered, they can realize surface-bound waves (\textit{e}.\textit{g}., surface plasmon polaritons at a metal-dielectric interface, which we call ST-SPPs) that have the same unique characteristics of their freely propagating counterparts. However, because the spatio-temporal spectral structure of ST-SPPs is key to their propagation invariance on the metal surface, their excitation methodology must be considered carefully. We show here using finite-difference time-domain (FDTD) simulations that an appropriately synthesized ST wave packet in free space can be coupled to a ST-SPP via a single nano-scale slit inscribed in the metal surface. Our calculations confirm that this excitation methodology yields surface-bound ST-SPPs that are localized in all dimensions (and can thus be considered as plasmonic `bullets'), which travel rigidly at the metal-dielectric interface without diffraction or dispersion at a tunable group velocity.
\end{abstract}

\maketitle

\section{Introduction}
Surface plasmon polaritons (SPPs) are surface-bound waves comprised of collective oscillations of free electrons at a metal-insulator interface \cite{Otto68ZP,Economou69PR,Kretschman71ZP,Raether88Book,Maier07Book,Zhang2012}. Their strong sub-wavelength spatial localization at the interface is the enabling feature for a host of applications ranging from super-resolution imaging \cite{Fang05Science,Kawata08NP,Willets17CR,Lee20NL} to sensing \cite{Anker08NM,Homola08CR}. However, similarly strong spatial localization of the SPP in its \textit{transverse} dimension parallel to the interface is thwarted by rapid diffractive spreading. Transverse SPP localization necessitates additional structuring of the metal surface (\textit{e}.\textit{g}., a waveguiding structure \cite{Bozhevolnyi06Nature,Oulton08NP,Fang15Light}), which usually increases SPP dispersion and thus diminishes the possibility of axial localization \cite{Pitarke,Kubo07NL}. Producing SPPs that are localized in all dimensions -- plasmonic bullets -- \textit{without} modifying the metal surface would be useful in sensing and imaging applications in which spatial and temporal localization need to be preserved over an extended distance.

A potential approach to overcoming transverse diffraction along the metal surface is to make use of particular spatial field profiles known as `diffraction-free' beams in the context of freely propagating monochromatic fields; \textit{e}.\textit{g}., Bessel \cite{Durnin87PRL} and Mathieu \cite{Gutierrez03AJP} beams (see also \cite{Levy16PO}). However, these field structures display non-diffracting behavior only when \textit{both} transverse dimensions are included. In contrast, SPPs are surface-bound waves in which only \textit{one} transverse dimension is available \cite{Wang18OE}. In such a configuration, the only potential diffraction-free field structures are the cosine wave (which is not localized) and the Airy beam \cite{Siviloglou07OL} (which travels along a parabolic trajectory rather than a straight line). Both of these special cases have been realized in plasmonic systems \cite{Salandrino10OL,Minovich11PRL,Zhang11OL,Li11PRL,Lin12PRL}. It remains however, to realize a SPP that is strongly localized in all three dimensions that also travels in a straight line. 

It has been recently shown that propagation invariance of pulsed optical fields traveling in a straight line can be realized in one transverse dimension by exploiting a precise form of space-time coupling \cite{Yessenov19OPN,Yessenov22AOP}. In these so-called `space-time' (ST) wave packets, each optical frequency undergirding the transverse beam profile is associated with a single temporal frequency (or wavelength) underlying the pulse profile \cite{Kondakci16OE,Parker16OE,Wong17ACSP2,Porras17OL,Efremidis17OL,Kondakci17NP}. As a consequence of this tight spatio-temporal spectral association, the ST wave packet propagates diffraction-free and dispersion-free in free space \cite{Bhaduri19OL}, non-dispersive dielectrics \cite{Bhaduri19Optica}, and dispersive media \cite{Hall22LPR}. Furthermore, ST wave packets exhibit a variety of unique behaviors; \textit{e}.\textit{g}., readily tunable group velocity \cite{Salo01JOA,Wong17ACSP2,Kondakci19NC}, anomalous refraction \cite{Bhaduri20NP,Motz21OL}, self-healing \cite{Kondakci18OL}, arbitrary axial acceleration \cite{Clerici08OE,Lukner09OE,Li20CP,Yessenov20PRL2,Li20SR,Li21CP,Hall22OLaccel}, new ST Talbot effects \cite{Hall21APLP}, among others \cite{Yessenov22AOP}. For our purposes here, the key property of ST wave packets that enable them to contribute to the quest for propagation-invariant SPPs is that their characteristics are retained when only one transverse coordinate is utilized \cite{Kondakci16OE,Kondakci17NP}. Indeed, a recent theoretical study \cite{Schepler20ACSP} has shown that such ST-SPPs are propagation invariant at metal-dielectric interfaces independently of the intrinsic SPP dispersion; their group velocity can be tuned above or below the SPP group velocity; and the ST-SPP can be localized in all dimensions to sub-wavelength size and yet not undergo diffractive spreading over the typical propagation distances dictated by ohmic losses in the metal. These characteristics indicate that ST-SPPs may usher in a unique platform for studying novel plasmonic phenomena.

The theoretical study of ST-SPPs \cite{Schepler20ACSP} left open the question regarding launching them. In general, free optical fields can excite SPPs by irradiating SPP-coupling structures; \textit{e}.\textit{g}., nano-scaled grooves, steps, and ridges \cite{Radko08ProcSPIE,Lalanne,Zhang13JPCC,Yang,Zayats, Ichiji22NanoP}. However, a fundamental difficulty remains in the case of ST-SPPs because conventional approaches do not readily lend themselves to the task of coupling a ST wave packet from free space to a ST-SPP; \textit{e}.\textit{g}., a grating will modify the spatio-temporal spectral structure of the incident field. Nevertheless, a recent experimental study demonstrated that a nanoslit inscribed in the metal surface can indeed couple a free broadband pulsed field associated with a pair of symmetric spatial frequencies, leading to a striped ST-SPP \cite{Ichiji22ACSP}. Such a nanoslit SPP-coupler has been previously exploited in coupling broadband optical pulses to conventional SPPs \cite{Kubo07NL,Zhang11PRB,Zhang13JPCC}. The question remains whether such a nanoslit-coupler can couple a spatio-temporally structured ST wave packet from free space to a surface bound ST-SPP \cite{DioufAO}.

In this paper, we carry out finite-difference time-domain (FDTD) simulations of the coupling of freely propagating ST wave packets to surface-bound SP-SPPs via a nanoslit in a metal surface \cite{Ichiji19OE}. Our simulations reveal the following findings: (1) By exciting the nanoslit SPP-coupler with individual monochromatic plane waves while tuning their wavelength and incident angle, we find only minute changes in the relative coupling efficiencies over the entire spectral and angular spans of interest. This indicates that the spectral amplitudes of the incident plane-wave components underlying an incident ST wave packet will be preserved in the excited ST-SPP. (2) We find that refraction at the interface between free space and the metal necessitates pre-compensation by modifying the spatio-temporal spectral structure of the incident ST wave packet. (3) By utilizing the appropriately prepared ST wave packet in free space, the nanoslit successfully excites propagation-invariant ST-SPPs at the metal interface with the targeted group velocity. (4) By adding a thin polymer layer on the metal surface to increase the SPP dispersion, we confirm that the nanoslit couples the free ST wave packet to a dispersion-free ST-SPP at the sought-after group velocity. Our results therefore indicate that a nanoslit SPP-coupler can be a simple yet versatile strategy for exciting ST-SPPs, studying their fascinating characteristics, and exploring their potential applications.

\begin{figure*}[t!]
  \begin{center}
  \includegraphics[width=17.2cm]{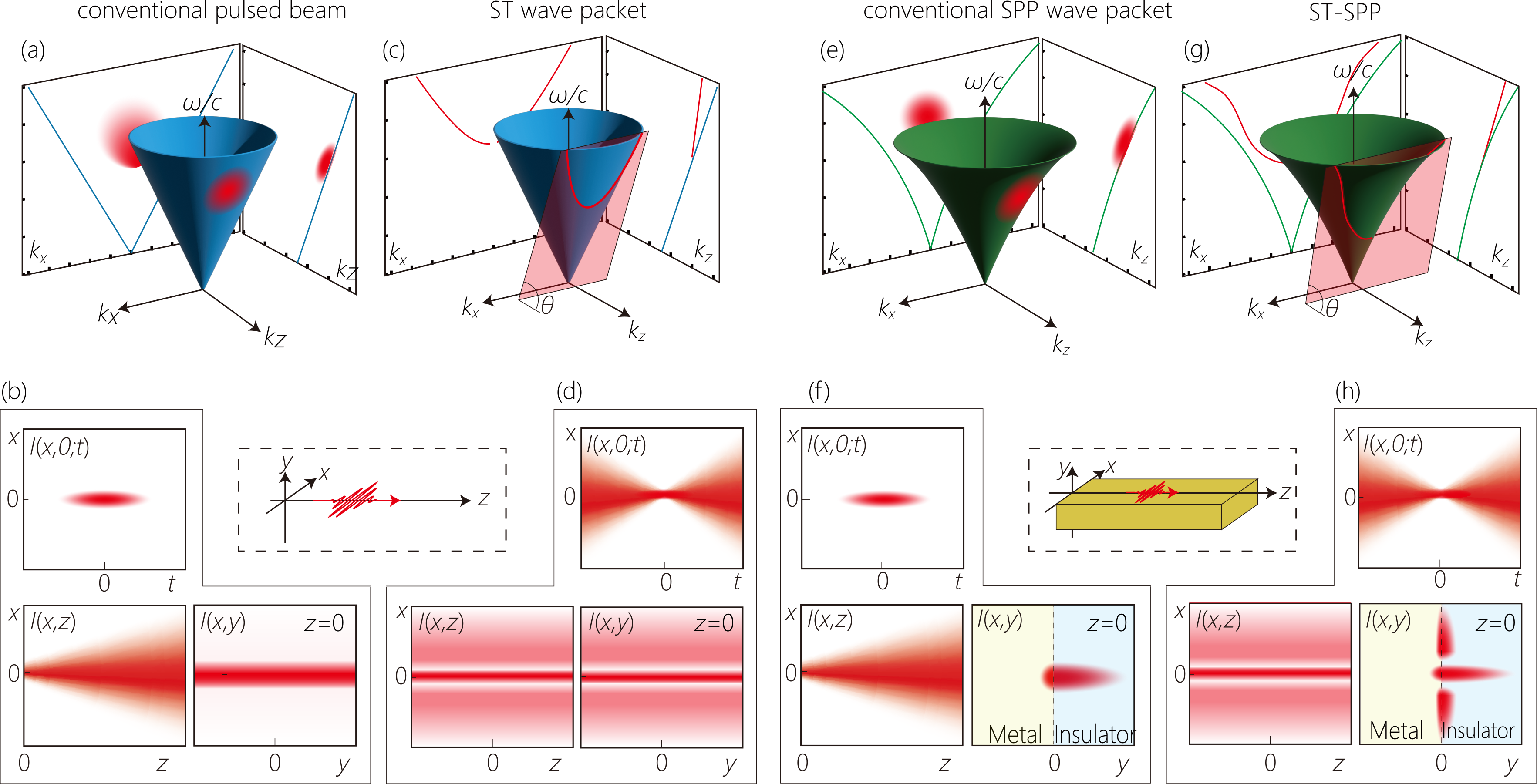}
  \end{center}
  \caption{(a) The spectral support for a conventional pulsed beam in free space on the surface of the light-cone $k_{x}^{2}+k_{z}^{2}\!=\!(\tfrac{\omega}{c})^{2}$ in $(k_{x},k_{z},\tfrac{\omega}{c})$-space. (b) Plots of three representations of the field in physical space: the initial spatio-temporal intensity profile $I(x,z=0;t)$; the time-averaged axial evolution $I(x,z)\!=\!\int\!dt\,|E(x,z;t)|^{2}$; and the time-averaged intensity in the transverse plane $I(x,y)$ at $z=0$. (c,d) Same as (a,b) but for a superluminal ST wave packet ($\theta\!>\!45^{\circ}$). The inset between (a,b) and (c,d) depicts the coordinate system in physical space. (e) Spectral support domain for a conventional pulsed SPP on the surface of the SPP light-cone $k_{x}^{2}+k_{z}^{2}\!=\!k_{\mathrm{SPP}}^{2}$. (f) The same intensity profiles as in (b) and (d). Here the intensity decays along the $y$-axis away from the metal-dielectric interface. (g,h) Same as (e,f) but for a surface-bound ST-SPP. The inset between (e,f) and (g,h) depicts the coordinate system along the metal surface in physical space.}
  \label{Fig:Concept}
\end{figure*}

\section{Theoretical background}
\subsection{Theory of ST wave packets in free space}

We first briefly describe the concept of ST wave packets in free space. We select $z$ as the direction of axial propagation, we hold the field uniform along the transverse coordinate $y$ (in anticipation of the implementation of ST-SPPs), so that the field is localized along only the other transverse coordinate $x$. In general, the field $E(x,z;t)\!=\!e^{i(k_{\mathrm{o}}z-\omega_{\mathrm{o}}t)}\psi(x,z;t)$ for a scalar pulsed beam (or wave packet) has an envelope $\psi$ given by \cite{SalehBook07}: 
\begin{equation}
\psi(x,z;t)=\iint\!dk_{x}d\Omega\;\widetilde{\psi}(k_{x},\Omega)e^{i(k_{x}x+(k_{z}-k_{\mathrm{o}})z-\Omega t)},
\end{equation}
where $\omega$ is the temporal angular frequency, $c$ is the speed of light in vacuum, $\omega_{\mathrm{o}}$ is the carrier frequency, $k_{\mathrm{o}}\!=\!\tfrac{\omega_{\mathrm{o}}}{c}$ is the corresponding wave number, $\Omega\!=\!\omega-\omega_{\mathrm{o}}$ is the frequency measured with respect to $\omega_{\mathrm{o}}$, $k_{x}$ and $k_{z}$ are the real components of the wave vector along the transverse and axial coordinates $x$ and $z$, respectively, which satisfy the dispersion relationship $k_{x}^{2}+k_{z}^{2}\!=\!(\tfrac{\omega}{c})^2$, and the spatio-temporal spectrum $\widetilde{\psi}(k_{x},\Omega)$ is the two-dimensional Fourier transform of $\psi(x,0;t)$. For a generic pulsed beam, the spatio-temporal spectrum is represented in $(k_{x},k_{z},\tfrac{\omega}{c})$- space by a 2D support domain on the surface of the light-cone $k_{x}^{2}+k_{z}^{2}\!=\!(\tfrac{\omega}{c})^{2}$, as shown in Fig.~\ref{Fig:Concept}(a). Such optical fields of course undergo diffractive spreading with propagation [Fig.~\ref{Fig:Concept}(b)], and are also deformed via space-time coupling in the case of spatially localized ultrashort pulses \cite{SalehBook07,Akturk10JO,Wikmark19PNAS,Leroux20OE,Jeandet22OE}.

In contrast, ST wave packets are a special class of \textit{propagation-invariant} pulsed beams whose spectral support on the surface of the light-cone is \textit{not} two-dimensional, but is instead restricted to a one-dimensional spectral trajectory \cite{Kondakci16OE,Kondakci17NP}; see Fig.~\ref{Fig:Concept}(c). Specifically, the spectral support is the intersection of the light-cone surface with a plane $\Omega\!=\!(k_{z}-k_{\mathrm{o}})c\tan{\theta}$, which is parallel to the $k_{x}$-axis and makes an angle $\theta$ (the spectral tilt angle) with the $k_{z}$-axis. Consequently, (1) each temporal frequency $\omega$ is associated with a particular spatial frequency $\pm k_{x}$; (2) the projection of the spectral support onto the $(k_{z},\tfrac{\omega}{c})$-plane is a straight line ($d\Omega/dk_{z}\!=\! \widetilde{v} \!=\!c\tan{\theta}$, a constant independent of $\omega$), and thus all higher order dispersions are zero; and (3) the spectral projection onto the $(k_{x},\tfrac{\omega}{c})$-plane is approximated by a parabola [Fig.~\ref{Fig:Concept}(c)]. Within this formulation, the envelope of the ST wave packet is:
\begin{equation}
\psi(x,z;t)\!=\!\!\int\!d\Omega\;\widetilde{\psi}(\Omega)e^{ik_{x}x}e^{-i\Omega(t-\tfrac{z}{\widetilde{v}})}\!=\!\psi(x,0;t-\tfrac{z}{\widetilde{v}}),
\end{equation}
where the transverse wave number (or spatial frequency) is $k_{x}(\Omega)\!=\!\sqrt{(k_{\mathrm{o}}+\Omega/c)^2-(k_{\mathrm{o}}+\Omega/\widetilde{v})^{2}}$, and we have made use of the substitution $k_{z}\!=\!k_{\mathrm{o}}+\tfrac{\Omega}{\widetilde{v}}$.

The ST wave packet travels rigidly in free space (dispersion-free and diffraction-free) at a group velocity $\widetilde{v}\!=\!c\tan{\theta}$ that depends only on the spectral tilt angle $\theta$. The ST wave packet is subluminal in free space $\widetilde{v}\!<\!c$ when $\theta\!<\!45^{\circ}$ and superluminal $\widetilde{v}\!>\!c$ when $\theta\!>\!45^{\circ}$. The spatio-temporal profile $I(x,z;t)\!=\!|E(x,z;t)|^{2}$ at any axial plane $z$ is usually X-shaped (see \cite{Kondakci18PRL,Wong21OE} for exceptions), and the time-averaged intensity $I(x,z)\!=\!\int\!dt\,I(x,z;t)$ is independent of $z$ [Fig.~\ref{Fig:Concept}(d)]. See Refs.~\cite{Bhaduri18OE,Yessenov19OE,Bhaduri19OL,Kondakci19OL} for a discussion of the upper limit on the propagation distance of ST wave packets as dictated by the physical constraints in a realistic experiment. Critically, ST wave packets are propagation invariant even when their transverse profile is restricted to one dimension ($x$ alone), so that ST wave packets may couple to propagation-invariant surface-bound SPPs.

\subsection{Theory of ST-SPPs at a metal-dielectric interface}

The dispersion relationship for a SPP at a metal-dielectric interface is $k_{\mathrm{SPP}}\!=\! (\frac{\omega}{c})\sqrt{\tfrac{\epsilon_{\mathrm{m}}(\omega)\epsilon_{\mathrm{d}}}{\epsilon_{\mathrm{m}}(\omega)+\epsilon_{\mathrm{d}}}}$, where $\epsilon_{\mathrm{d}}$ and $\epsilon_{\mathrm{m}}$ are the relative permittivities of the dielectric and metal, respectively. For a metal surface in free space, which is the main subject in this paper, $\epsilon_{\mathrm{d}}$ is assumed to be unity independent of $\omega$. The field is bound to the metal-dielectric interface and decays away from it along $y$ in both media. When the SPP has a finite transverse spatial width (SPP beam) and a finite temporal linewidth (SPP pulse), the spectral support of the resulting wave packet on the surface of the SPP light-cone $k_{x}^{2}+k_{z}^{2}\!=\!k_{\mathrm{SPP}}^{2}$ is a 2D domain [Fig.~\ref{Fig:Concept}(e)]; the SPP field is now localized along $x$ and $t$. However, the SPP transverse spatial profile spreads diffractively with propagation along $z$ [Fig.~\ref{Fig:Concept}(f)]. In addition the SPP temporal pulse profile undergoes dispersive spreading by virtue of the  group velocity dispersion (GVD) intrinsic to SPP propagation. Nevertheless, because the SPP propagation distance along the metal-dielectric interface is limited by ohmic losses, dispersive pulse broadening is relevant only for ultrashort pulses or for highly dispersive plasmonic systems.

\begin{figure*}[ht!]
  \begin{center}
  \includegraphics[width=16.8cm]{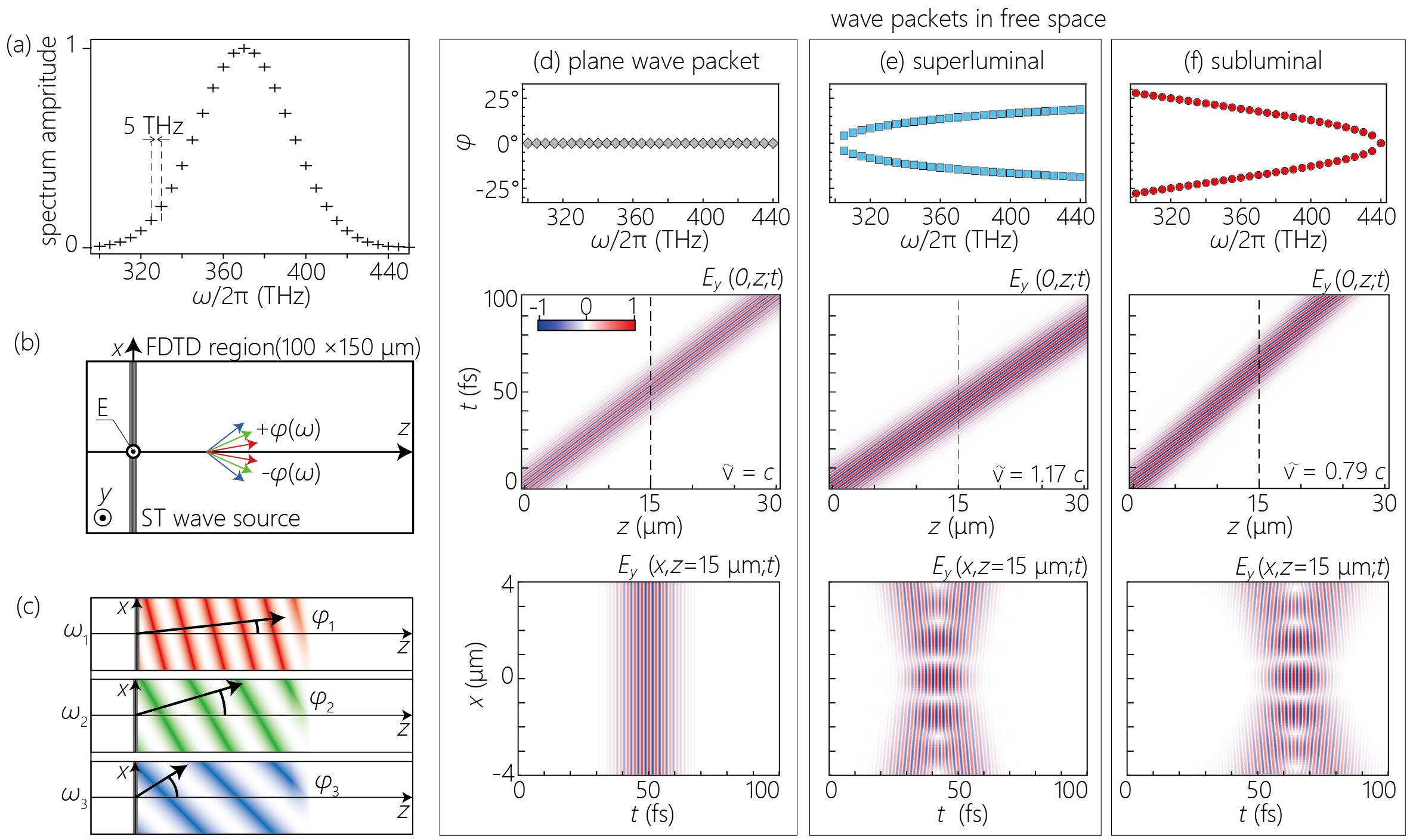}
  \end{center}
  \caption{(a) Spectrum used in the FDTD simulations. (b) Schematic of the FDTD model and simulation space. (c) Each frequency $\omega$ may be directed at a different angle $\varphi(\omega)$ with respect to the $z$-axis. Here, three frequencies $\omega_{1}$, $\omega_{2}$, and $\omega_{3}$ (depicted as different colors) propagate at different angles $\pm\varphi_{1}$, $\pm\varphi_{2}$,  and $\pm\varphi_{3}$, respectively. (d-f) Results of the FDTD simulations: (d) spatially plane wave packet; (e) superluminal ST wave packet ($\widetilde{v}\!=\!1.2c$); and (f) subluminal ST wave packet ($\widetilde{v}\!=\!0.8c$) -- all in free space. The first row shows the frequency-dependent propagation angle $\varphi(\omega)$; the second row shows the real part of the electric field at the beam center $E_{y}(0,z;t)$; and the third row depicts $E_{y}(x,z;t)$ at a fixed axial plane $z\!=\!15$~$\mathrm{\mu}$m (corresponding to the plane identified with the dashed vertical line in the second row).}
  \label{Fig:FDTDFreeSpace}
\end{figure*}

Both diffraction and dispersion that plague strongly localized SPPs can be thwarted by making use of ST-SPPs, which are the surface-wave analog of free-space ST wave packets [Fig.~\ref{Fig:Concept}(g,h)]. By restricting the spectral support of the SPP wave packet on the surface of the SPP light-cone to the 1D trajectory at its intersection with a plane $\Omega\!=\!(k_{z}-k_{\mathrm{o}}')c\tan{\theta}$, which is parallel to the $k_{x}$-axis and makes an angle $\theta$ with the $k_{z}$-axis [Fig.~\ref{Fig:Concept}(g)], we obtain a ST-SPP that propagates invariantly at a group velocity $\widetilde{v}\!=\!c\tan{\theta}$ \cite{Schepler20ACSP}. Here $\Omega\!=\!\omega-\omega_{\mathrm{o}}$, and $k_{\mathrm{o}}'$ is the wave number on the SPP light-line associated with $\omega\!=\!\omega_{\mathrm{o}}$; $k_{\mathrm{o}}'\!\neq\!\tfrac{\omega_{\mathrm{o}}}{c}$. The 1D spectral trajectory for a ST-SPP on the SPP light-cone surface can be obtained from the constraints $k_{z}(\omega)\!=\!k_{\mathrm{o}}'+\tfrac{\Omega}{\widetilde{v}}$ and $k_{x}(\omega)\!=\!\sqrt{k_{\mathrm{SPP}}^{2}-(k_{\mathrm{o}}'+\tfrac{\Omega}{\widetilde{v}})^{2}}$.

\section{FDTD simulations of free-space space-time wave packets}
\subsection{FDTD model for ST wave packets in free space}
We describe here our methodology for investigating the propagation of ST wave packets in free space using FDTD simulations (making use of the software package FDTD Solution, Ansys Lumerical). We carry out this step as a prelude to FDTD simulations of the coupling from free-space ST wave packets to a surface-bound ST-SPP. The spectral amplitudes at each frequency $\omega$ follow the Gaussian profile resulting from the Fourier transform of the electric field, $E(t)\!\propto\!e^{-(t/\Delta t)^{2}}\cos\omega_{\mathrm{o}}t$, where $\Delta t\!=\!10$~fs (the full width at half maximum (FWHM) of 16.7~fs) and $\tfrac{\omega_{\mathrm{o}}}{2\pi}\!=\!370$~THz [Fig.~\ref{Fig:FDTDFreeSpace}(a)]. We sample the spectrum at 5 THz intervals in the range of 300~THz to 440~THz. The overall layout of the geometry employed in the FDTD simulations is illustrated in Fig.~\ref{Fig:FDTDFreeSpace}(b). These wave sources (referred to hereafter as the `ST-wave source') are superposed in the FDTD simulation to reproduce the propagation dynamics of the target wave packet. Each frequency component $\omega$ is set up as a continuous wave source directed at an angle $\varphi(\omega)$ with respect to the $z$-axis [Fig.~\ref{Fig:FDTDFreeSpace}(c)]. We place all sources at $z\!=\!0$; periodic boundary conditions are applied along the $y$-axis; absorbing boundary conditions (perfect matched layers) to both ends of the $x$- and $z$-axes; and the calculation area along $x$ and $z$ is sufficiently large to avoid artifacts caused by waves reflected from the boundaries [Fig.~\ref{Fig:FDTDFreeSpace}(b)].
For a wave packet with the field temporally confined and the wavefront infinitely and uniformly spread in free space (referred to hereafter as `spatially plane wave packet'), all the frequency components travel in the same direction along the z-axis [Fig.~\ref{Fig:FDTDFreeSpace}(d)]. In the case of free-space ST wave packets [Fig.~\ref{Fig:FDTDFreeSpace}(e,f)], the angles of the individual continuous-wave sources are given by $\varphi(\omega)\!=\!\sin^{-1}(\tfrac{ck_{x}(\omega)}{\omega})$, which we plot in Fig.~\ref{Fig:FDTDFreeSpace}(e,f) for superluminal ($\widetilde{v}\!=\!1.2c$) and subluminal ($\widetilde{v}\!=\!0.8c$) ST wave packets.

\begin{figure*}[ht!]
  \begin{center}
  \includegraphics[width=16.8cm]{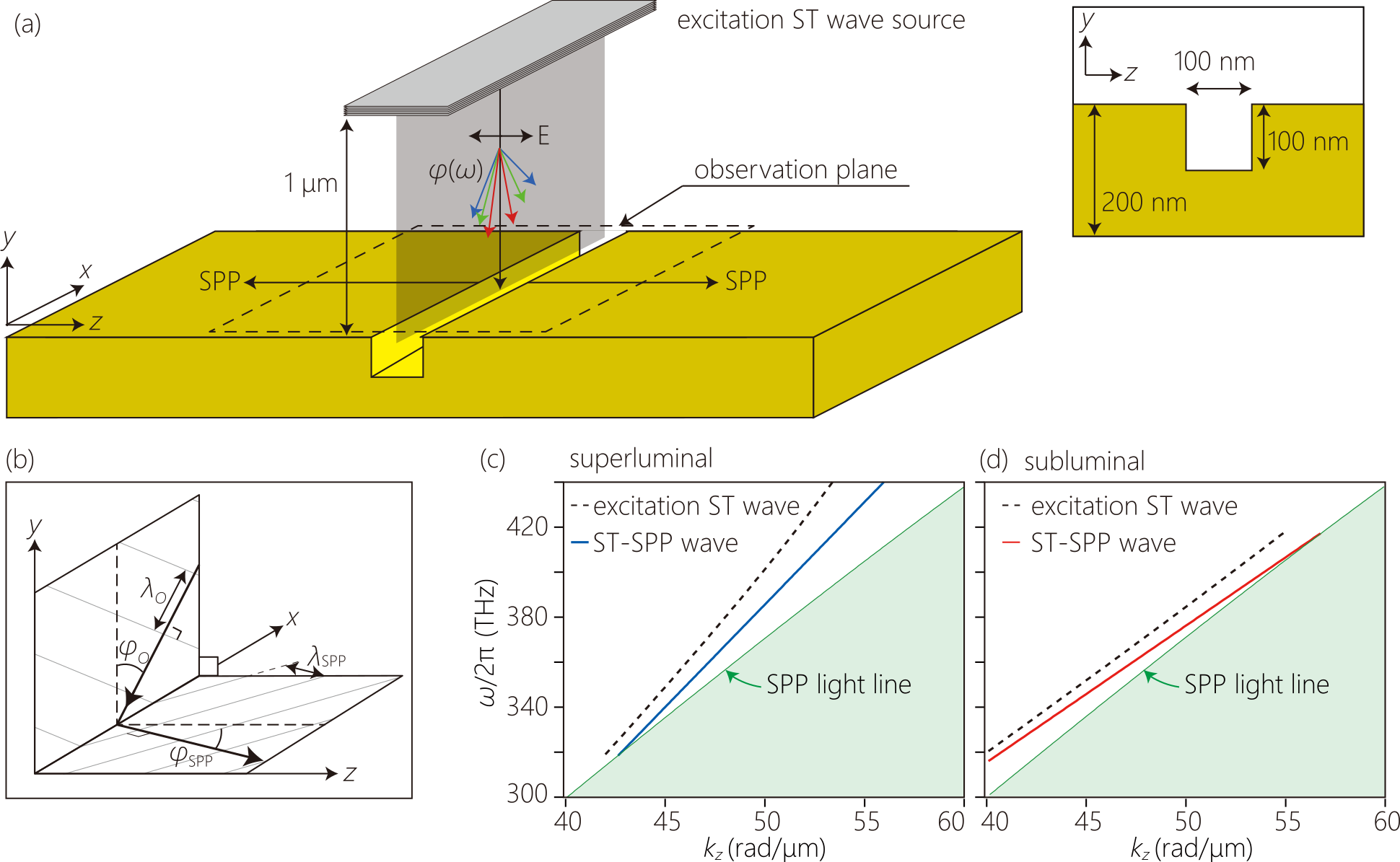}
  \end{center}
   \caption{(a) Schematic of the FDTD simulation model for a ST-SPP wave packet. Inset shows the dimensions of the nanoslit structure. (b) Excitation of SPP by obliquely incident on the excitation structure from free-space. (c) Projection onto the ($k_{z},\omega$)-plane of the spectral support domain for a superluminal ST-SPP ($\widetilde{v}\!=\!1.2c$) and the excitation ST wave packet, along with the SPP light-line. (d) Same as (c) for a subluminal ($\widetilde{v}\!=\!0.8c$) ST-SPP.}
  \label{Fig:FDTDSPP}
\end{figure*}

\subsection{Simulations of ST wave packets in free space}
The results of the FDTD calculations in free space are plotted in  Fig.~\ref{Fig:FDTDFreeSpace}(d-f). We first plot the axial evolution of the pulse profile Re$\{E_{y}(0,z;t)\}$ at the beam center $x\!=\!0$, with the axes selected such that the wave-packet peak traces a diagonal trajectory from the bottom-left corner to the top-right corner when $\widetilde{v}\!=\!c$ (a propagation distance of 30~$\mu$m in 100~fs). The spatially plane wave packet in Fig.~\ref{Fig:FDTDFreeSpace}(d) indeed traverses such a trajectory, thereby confirming that $\widetilde{v}\!=\!c$. In contrast, the corresponding calculations for superluminal and subluminal ST wave packets in Fig.~\ref{Fig:FDTDFreeSpace}(e) and Fig.~\ref{Fig:FDTDFreeSpace}(f), respectively, reveal different trajectory slopes with respect to the $z$-axis. The group velocities of the simulated ST wave packets estimated from their peaks are $\widetilde{v}\!\approx\!1.17c$ for the superluminal ST wave packet and $\widetilde{v}\!\approx\!0.79c$ for the subluminal ST wave packet, which are consistent with the target values.

We also plot in Fig.~\ref{Fig:FDTDFreeSpace}(d-f) the spatio-temporal profile of the electric field Re$\{E_{y}(x,z;t)\}$ at a fixed plane $z\!=\!15$~$\mu$m. In the case of the spatially plane wave packet, the spatial and temporal degrees of freedom are separable: the field is uniform along $x$ and Gaussian along $t$. In contrast, the superluminal and subluminal ST wave packets are X-shaped, with the superluminal wave packet arriving earlier with respect to the spatially plane wave packet, and the subluminal wave packet arriving later. Additionally, the X-shaped profile remains invariant across the entire observation plane (see Supplementary Movie1 and Movie2). From these simulations, we confirm that the FDTD model we have adopted for ST wave packets reproduces their expected propagation characteristics in agreement with previous theoretical and experimental results \cite{Yessenov22AOP}.

\section{FDTD simulations of a space-time SPP on a metal surface}

\subsection{FDTD model for ST-SPPs}

We next proceed to describing the FDTD model for coupling a free-space ST wave packet to a surface-bound ST-SPP at a gold (Au)-air interface [Fig.~\ref{Fig:FDTDSPP}(a)], which consists of a ST wave packet source, a nanoslit SPP coupler, and the Au surface. (The dispersive dielectric function of Au was determined according to Rakic \textit{et al.} \cite{Rakic98AO}) In this model, to excite a ST-SPP a free-space ST wave packet is incident perpendicularly on the structure designed to excite a ST-SPP (hereafter referred to as the excitation ST wave source). We place a single slit of depth 100 nm and width 100 nm on the Au surface as the SPP-coupler \cite{Kubo07NL,Zhang13JPCC,Zhang11PRB}; see Fig. 3(a) inset. When exciting SPPs at a specific wavelength, a grating structure is typically used. Here, we use a single nanoslit to preserve the spatio-temporal spectral structure associated with ST-SPPs. With the dimension of the nanoslit set here, the resonance frequencies of the slit do not overlap in the spectral range of the incident light.

Because the nanoslit has translational symmetry along the $x$-axis, the spatial frequency $k_{x}$ is conserved across the interface between free space and the metal surface. However, we must take into account the change in the angle $\varphi(\omega)$ resulting from the $k_{x}$-invariance. For example, a continuous wave source at a wavelength $\lambda_{\mathrm{o}}=\tfrac{2\pi}{k_{\mathrm{o}}}$ and an incident angle $\varphi_{\mathrm{o}}$ in free space is coupled to an SPP excited at the metal surface having a propagation angle $\varphi_{\mathrm{SPP}}$ as determined by $k_{\mathrm{o}}\sin\varphi_{\mathrm{o}}=k_{\mathrm{SPP}} \sin\varphi_{\mathrm{SPP}}$ [Fig.~\ref{Fig:FDTDSPP}(b)]. Therefore, to produce a propagation-invariant ST-SPP satisfying the conditions $k_{z}\!=\!k_{\mathrm{o}}'+\Omega/\widetilde{v}$ and $k_{x}\!=\!\sqrt{k_{\mathrm{SPP}}^{2}-(k_{\mathrm{o}}'+\Omega/\widetilde{v})^{2}}$, we must set the spatial frequency of the incident excitation $k_{x}^{(\mathrm{ex.})}$ equal to that of the spatial frequency of the ST-SPP $k_{x}$ at the same frequency $\omega$, $k_{x}^{(\mathrm{ex.})}(\omega)\!=\!k_{x}(\omega)$ to pre-compensate for the refractive change in angle. This is brought out in Fig.~\ref{Fig:FDTDSPP}(c,d) where we plot the spectral projections onto the $(k_{z},\omega)$-plane for a ST-SPP and the incident free-space ST wave packet excitation. The SPP light-line is of course curved because of SPP dispersion. The spectral projection of the ST-SPP is a straight line (indicating absence of dispersion of any order). However, the spectral projection of the excitation ST wave packet is curved, indicating that the free-space ST wave packet must be dispersive in order to excite a propagation-invariant ST-SPP. Therefore the ST wave packet is not propagation invariant here. The duration of the pulse extends while the propagation in the free space according to the dispersion adjunct for making the spectral projection of the ST-SPP straight \cite{Hall22LPR}. The magnitude of the broadening in the free space is, however, limited to a negligible level in the FDTD calculations due to the short distance from the plane wave source to the nanoslit. Note that in practical experiments with optics for synthesis of the ST wave packet, an additional system for dispersion compensation is necessary to compensate for the broadening in free space and obtain the designed pulse structure on metal surfaces. In our simulations, we launch a superluminal ST-SPP with target group velocity $\widetilde{v}\!=\!1.2c$ and a subluminal ST-SPP with the $\widetilde{v}\!=\!0.8c$.

\begin{figure}[ht!]
\centering
\includegraphics[width=8.6 cm]{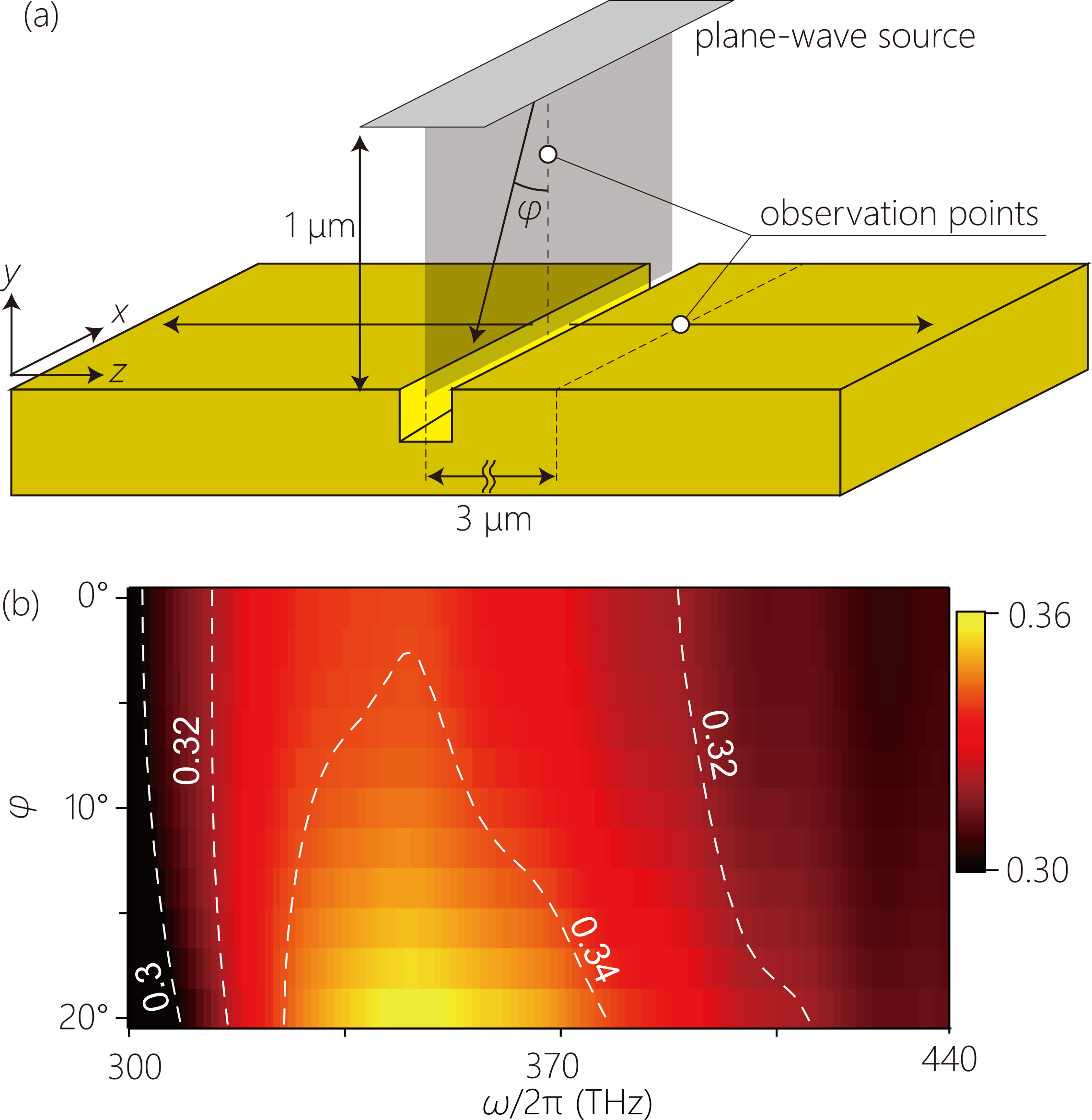}
\caption{(a) Schematic of the FDTD simulation model to evaluate the impact of the incident angle and wavelength on the relative excitation efficiency of the SPPs. (b) Plot of the spectral intensities of the SPPs with incident angle and frequency, normalized with respect to the spectral amplitudes of the excitation waves.}
\label{Fig:dependency}
\end{figure}

\subsection{Relative SPP coupling efficiencies}

A central goal in this study is to ascertain the utility of a nanoslit in coupling a free-space ST wave packet to a surface-bound ST-SPP. An important criterion for assessing this functionality is whether all the plane-wave components in the free-space ST wave packet couple with approximately equal efficiency to the ST-SPP across the spectral bandwidth and angular range of interest. We investigate this question by directing individual plane waves from free space at frequency $\omega$ and incident angle $\varphi$, as illustrated in Fig.~\ref{Fig:dependency}(a). Two observation points are set in free space (500~nm down from the plane wave source) and on the Au surface (3~$\mu$m to the right of the nanoslit) to record the temporal electric field waveforms of the excitation wave and SPP wave packet, respectively. 

The results of the FDTD calculations of the spectral amplitudes of the excited SPPs (normalized with respect to the spectral amplitudes of the free-space excitation) are plotted in Fig.~\ref{Fig:dependency}(b). We find that there is only a small variation in the relative coupling efficiency across the spectral and angular spans of interest. The insensitivity to the  frequency makes the spectral shape of the ST-SPP almost identical to the excitation light in free space. In addition, the nearly uniform, structure-free coupling efficiency suggests that the resonance modes of the nanoslit are not in the frequency range of interest as intended. The phase relation of the frequency components formed in the ST wave packet is preserved through the coupling. However, notable decrease was observed above 500~THz due to absorption in Au. With this potential obstacle now cleared, we proceed to the results of FDTD simulations of coupling ST wave packets to ST-SPPs.

\begin{figure*}[ht!]
  \begin{center}
  \includegraphics[width=17.2cm]{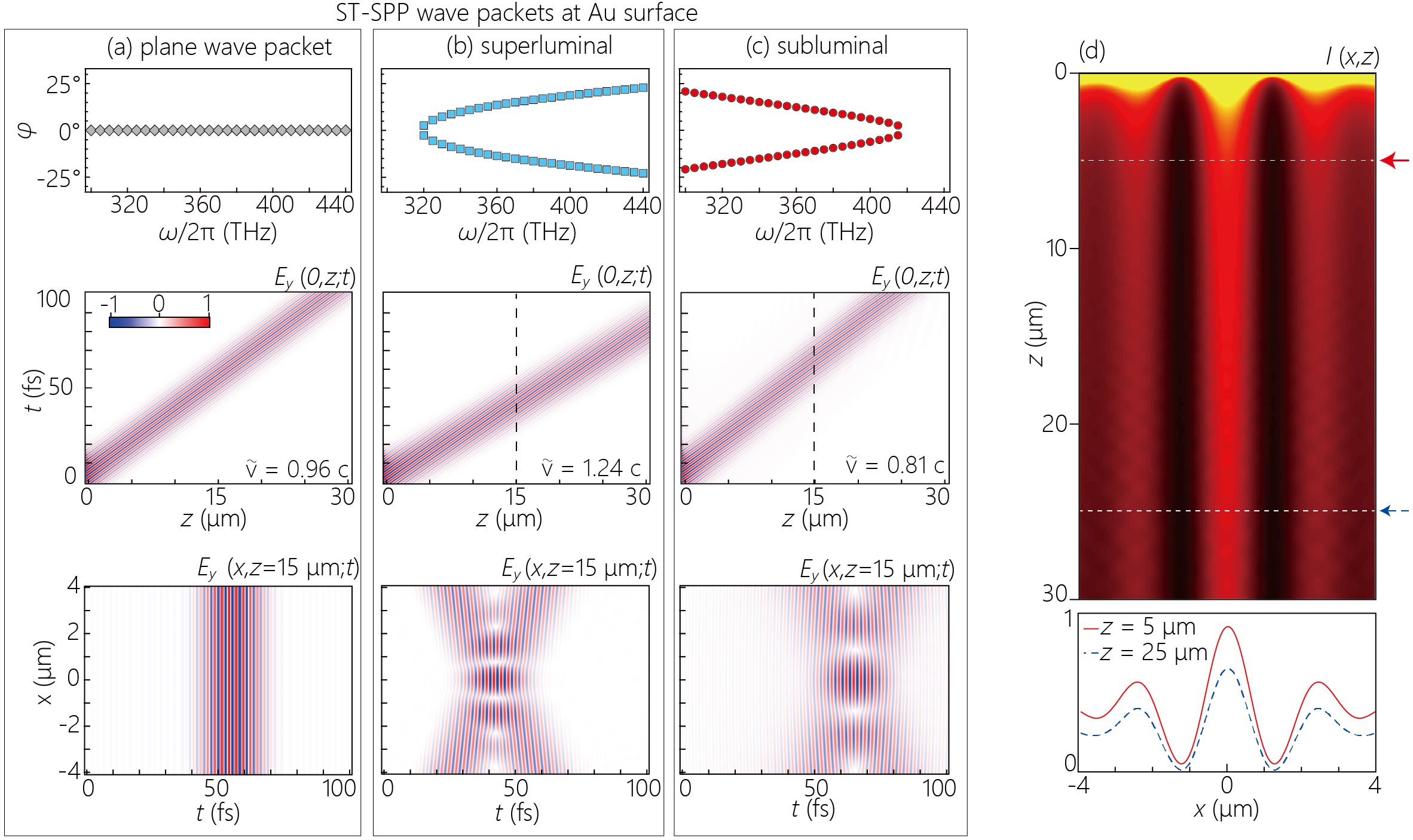}
  \end{center}
\caption{(a-c) Results of FDTD simulations for (a) a spatially plane SPP wave packet, (b) a superluminal ST-SPP with $\widetilde{v}\!=\!1.2c$, and (c) a subluminal ST-SPP with $\widetilde{v}\!=\!0.8c$. The rows are arranged in the same fashion as Fig.~\ref{Fig:FDTDFreeSpace}(d-f) (d) Time-averaged intensity $I(x,z)$ for the superluminal ST-SPP. The intensity at two axial planes $z\!=\!5$~$\mu$m and $25$~$\mu$m are plotted at the bottom of the panel for comparison.}
\label{Fig:Result}
\end{figure*}

\subsection{Simulation results for ST-SPPs on the metal surface}

We plot in Fig.~\ref{Fig:Result}(a-c) the simulation results for the SPP model in Fig.~\ref{Fig:FDTDSPP}(a). A small temporal delay ($\sim5$~fs in the cases considered here ; see $E_{y}(0,z;t)$ plots in Fig.~\ref{Fig:Result}(a-c)) accrues in the excited SPPs because of the distance between the wave source and the nanoslit. By tracing the trajectory of the field's peak in the $(z,t)$ plane, we confirm that the group velocities of the excited SPPs on the Au surface match the target values. The group velocity of the spatially plane SPP wave packet in Fig.~\ref{Fig:Result}(a) is slightly lower than $c$ due to SPP dispersion ($\widetilde{v}_{\mathrm{SPP}}\!=\!0.96c$). For the superluminal and subluminal ST-SPPs, we find $\widetilde{v}=1.24c$ and $\widetilde{v}=0.81c$, which are consistent with the target values. The X-shaped field distributions of the ST-SPPs in Fig.~\ref{Fig:Result}(b,c) ($E_{y}(x,z=15\mu$m$;t)$) differ only slightly from their free-space counterparts (in Fig.~\ref{Fig:FDTDFreeSpace}(e,f)) due to SPP dispersion (see Supplementary Movie3, Movie4).

The diffraction-free propagation of the ST-SPPs launched by the nanoslit at $z=0$ is confirmed from the time-averaged intensity $I(x,z)\!=\!\int\!dt\;|E_{y}(x,z;t)|^{2}$ [Fig.~\ref{Fig:Result}(d)]. The attenuation of the intensity with propagation along $z$ is caused by the imaginary part of the dispersion relation of the SPP on the Au surface \cite{Schepler20ACSP}. The FDTD simulations have thus verified that the nanoslit structure couples free-space ST wave packets to ST-SPPs at the target group velocity.

\subsection{Large-dispersion model}

\begin{figure}[t]
\centering
\includegraphics[width=8.6 cm]{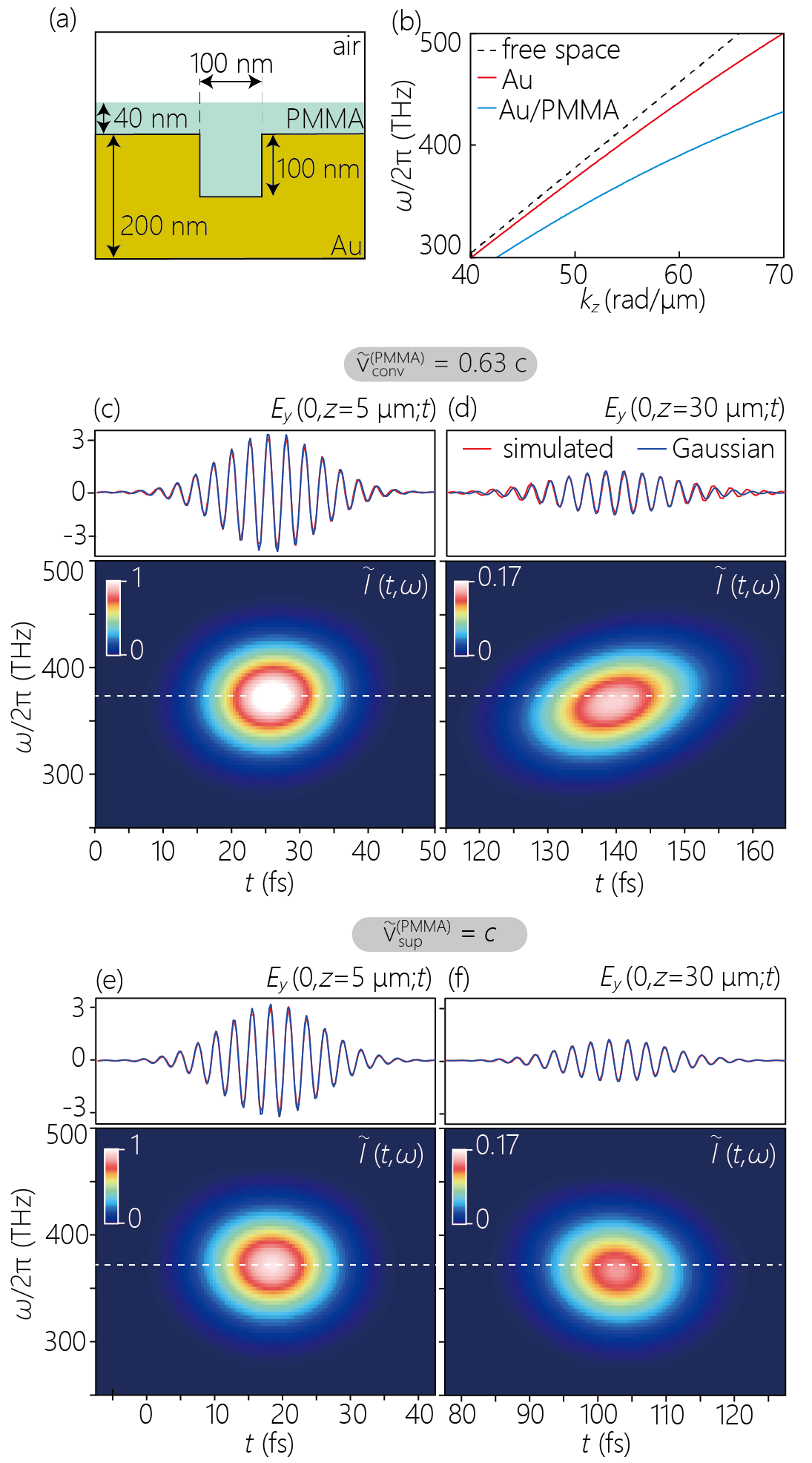}
\caption{(a) Schematic of the geometry of the nanoslit for the `large-dispersion' model. (b) Dispersion curves (light-lines) for propagation in free space; along the Au-air interface; and along the Au-PMMA interface. (c,d) Plots of the real part of the field $E_{y}(0,z;t)$ temporal waveform for a conventional SPP wave packet, and the time-frequency spectrograms $\widetilde{I}(t,\omega)$ at $z=5$ and 30~$\mu$m. (e,f) Same plots as shown in (c,d) for a superluminal ST-SPP wave packet. The dashed white lines indicate the central frequency of the excited wave.}
\label{Fig:PMMA}
\end{figure}

The short propagation distances typically associated with SPPs preclude the observation of GVD-related effects. Thus, only the diffraction-free propagation of ST-SPPs is prominent, whereas their dispersion-free propagation is not. To bring this aspect of ST-SPPs out, we modify the SPP model shown in Fig.~\ref{Fig:FDTDSPP}(a) to introduce higher dispersion into the SPP propagation, referred henceforth as the `large-dispersion' model. This is accomplished by adding a 40-nm-thick poly(methyl methacrylate) (PMMA; relative dielectric constant $\epsilon\!=\!2.3$) layer on the previous model [Fig.~\ref{Fig:PMMA}(a)]. The metal surface covered by the thin insulator film has been shown to have large dispersion \cite{Lemke, Bozhevolnyi}. We calculate the dispersion relation for the Au-PMMA system, $k_{\mathrm{PMMA}}(\omega)$ using the four-layer model proposed by Pockrand \cite{pockrand}; see Fig.~\ref{Fig:PMMA}(b) for the calculated dispersion curves of the bare Au surface and the Au-PMMA system. This model allows us to address two critical questions: (a) Can the nanoslit couple a free-space ST wave packet to a ST-SPP in a metal-insulator plasmonic system? (2) Can the surface-bound ST-SPP propagate dispersion-free in presence of large GVD?

In the FDTD simulations, we examine the excitation of a spatially plane SPP wave packet and a ST-SPP with a target superluminal group velocity of $\widetilde{v}\!=\!c$. We plot in Fig.~\ref{Fig:PMMA}(c,d) the temporal waveforms of the real part of the SPP electric field component $E_{y}$ at $z\!=\!5$ and 30~$\mathrm{\mu}$m in the spatially plane SPP wave packet model. By tracing the motion of the peak of the SPP wave packet, we estimate its group velocity as $\widetilde{v}_{\mathrm{PMMA}}\approx0.63c$; in agreement with the value expected from the dispersion curve plotted in Fig.~\ref{Fig:PMMA}(b). Whereas the waveform at $z=5$~$\mathrm{\mu}$m is well-fitted by a Gaussian-sinusoid function, that at $z\!=\!30$~$\mathrm{\mu}$m deviates from the fitting curve at both the head and tail of the wave packet. The FWHM of the temporal waves calculated from least-square fitting are 17.1 and 20.6~fs, respectively, thus indicating the onset of temporal dispersive broadening of the SPP wave packet. The associated chirp of the SPP wave packet caused by dispersion is revealed in the time-frequency spectrogram (calculated from the temporal waveform using the Wigner distribution function \cite{Ichiji19OE}), where an up-chirping in the carrier frequency is clear from the stretched distribution contour from the lower left to the upper right. 

In contrast, the ST-SPP [Fig.~\ref{Fig:PMMA}(e,f)] does \textit{not} suffer the deleterious impact of SPP dispersion. Instead, the ST-SPP is propagation invariant in both space and time. The temporal waveforms at $z\!=\!5$ and $30$~$\mathrm{\mu}$m are both well-fitted to Gaussian-sinusoidal functions with the FWHM of 17.2 and 17.8~fs, respectively. The spectrograms of the ST-SPP wave packets are nearly isotropic with no visible chirping. Moreover, the estimated group velocity of the ST-SPP is $\widetilde{v}\!\approx\!c$. The only change observed in the ST-SPP with propagation is an overall attenuation in intensity and slight red-shift at $z\!=\!30$~$\mathrm{\mu}$m (both shared with the plane SPP) caused by the dispersive propagation losses. In the frequency range of the ST-SPP, the imaginary part of the SPP dispersion relation gradually increases as the frequency increases, providing larger damping for higher frequencies. This feature induces a slight redshift in the center frequency of the ST-SPP  [Fig.~\ref{Fig:PMMA}(f)]. Nevertheless, the ST-SPP maintains the regulated time width and the spatial distribution. These results suggest that irradiating a single nanoslit with an excitation ST wave packet can help to excite ST-SPPs in complex plasmonic structures, such as a metal coated by an insulator layer or other systems featuring large dispersion.

\section{conclusion}

In conclusion, making use of FDTD simulations, we have investigated the potential for exciting surface-bound SPPs by coupling free-space ST wave packets to a metal surface via a nanoslit. These new plasmonic field structures, which we have called ST-SPPs, are propagation-invariant pulsed SPPs that are strongly localized in all dimensions, and yet propagate along the metal interface without diffraction or dispersion -- and thus can be considered plasmonic bullets. We demonstrate that by starting with freely propagating ST wave packets, in which each frequency travels at a prescribed angle with respect to the propagation axis, ST-SPPs are excited at a metal-dielectric interface via a single nano-slit introduced into the metal surface. Crucially, we have confirmed that the target group velocity of the ST-SPP is realized using this excitation methodology, and only small variations are found in the relative coupling coefficients of plane waves over the spans of frequencies and incident angles of interest. Furthermore, the resistance of ST-SPPs to dispersion was confirmed in a highly dispersive plasmonic system consisting of a thin polymer layer on top of the metal surface. From these FDTD simulations, we conclude that ST-SPPs may indeed be excited at a metal interface via a nanoslit irradiated with a free-space ST wave packet. FDTD simulations appear to be a useful calculation method for studying the excitation of ST-SPPs and examining their propagation characteristics in plasmonic structures. Moreover, these results indicate that the recent rapid progress in the study of structured light may also have impact on plasmonic wave packets \cite{Brongersma,Jolly20JOpt,Forbes21NatP,Piccardo22JOpt,Yessenov22AOP}.

\begin{acknowledgments}
This work was supported by the JSPS KAKENHI (JP16823280, JP18967972, JP20J21825); MEXT Q-LEAP ATTO (JPMXS0118068681); and by the U.S. Office of Naval Research (ONR) under contracts N00014-17-1-2458 and N00014-20-1-2789.
\end{acknowledgments}

\bibliographystyle{apsrev4-1}
\bibliography{sorsamp}

\end{document}